\documentclass[twocolumn,showpacs,preprintnumbers,amsmath,amssymb]{revtex4}
\usepackage{graphicx}
\usepackage{bm}

\begin{document}
\preprint{arXiv:hep-ph/0410155}
\title{Determining the Sign of the {\boldmath $b \to s \gamma$} Amplitude}
\author{Paolo Gambino}
\affiliation{INFN, Torino and Dipartamento di Fisica Teorica, 
                   Universit\`a di Torino, 10125 Torino, Italy}
\author{Ulrich Haisch}
\affiliation{Theoretical Physics Department, Fermilab, Batavia, Illinois 60510, USA}
\author{Miko{\l}aj Misiak}
\affiliation{Institute of Theoretical Physics, Warsaw University, 00-681 Warsaw, Poland}

\begin{abstract}
The latest Belle and BaBar measurements of the inclusive $\bar{B} \to X_s l^+
l^-$ branching ratio have smaller errors and lower central values than the
previous ones. We point out that these results indicate that the sign of the
$b \to s \gamma$ amplitude is the same as in the SM. This underscores the
importance of $\bar{B} \to X_s l^+ l^-$ in searches for new physics, and may
be relevant for neutralino dark matter analyses within the MSSM.
\end{abstract}

                             
\maketitle

The branching ratio of the inclusive radiative $B$-decay is one of the most
important constraints for a number of new physics models, because it is
accurately measured and its theoretical determination is rather clean.  The
present world average ${\cal B}(\bar B\to X_s \gamma)= (3.52 \pm 0.30)\times
10^{-4}$ \cite{hfag} agrees very well with the Standard Model (SM) prediction
${\cal B}(\bar B\to X_s \gamma)_{\rm SM}= (3.70 \pm 0.30) \times 10^{-4}$~\cite{bsgSM}.  
A well-known way to avoid this constraint without excluding large new physics
effects consists in having new physics contributions that approximately
reverse the sign of the amplitude $A(b \to s \gamma)$ with respect to the SM
and leave ${\cal B}(\bar{B} \to X_s \gamma)$ unaltered within experimental and
theoretical uncertainties.  Several authors pointed out that even a rather
rough measurement of the inclusive $\bar{B} \to X_s l^+ l^-$ branching ratio
could provide information on the sign of $A(b \to s
\gamma)$~\cite{Ali:1994bf}.

Other observables that are sensitive to the sign of $A(b \to s \gamma)$ are
the forward-backward and energy asymmetries in inclusive and exclusive $b \to
s l^+ l^-$ decays~\cite{Ali:1994bf, Ali:1991is}. Very recently, the first
measurement of the forward-backward asymmetry in $B \to K^{(*)} l^+ l^-$ was
announced by the Belle Collaboration~\cite{Abe:2004ir}. Within the limited
statistical accuracy, however, the results were found to be consistent with
both the SM and the ``wrong-sign'' $A(b \to s \gamma)$ case.

The purpose of this Letter is to point out that the present measurements of
${\cal B}(\bar{B} \to X_s l^+ l^-)$ already indicate that the sign of $A(b \to
s \gamma)$ is unlikely to be different from that in the SM.  The experimental
results that we consider are summarized in Tab.~\ref{tab:exp}.

The results in Tab.~\ref{tab:exp} are averaged over muons and electrons.  The
first range of the dilepton mass squared $q^2$ corresponds to the whole
available phase-space for  $l=\mu$, but includes a cut for $l=e$. Moreover,
the intermediate $\psi$ and $\psi'$ are treated as background, and a Monte
Carlo simulation based on perturbative calculations is applied for the
unmeasured part of the $q^2$-spectrum that hides under the huge $\psi$ and
$\psi'$ peaks (see Refs.~\cite{Abe:2004sg,Aubert:2004it} for more details). In
the second range of $q^2$ in Tab.~\ref{tab:exp}, theoretical uncertainties
are smaller than in the first case (see below), but the experimental errors
are larger due to lower statistics.  As we shall see, the analyses in both
regions lead to similar conclusions concerning the sign of $A(b \to s
\gamma)$.
\begin{table}[h]
\caption{\sf  
Measurements of ${\cal B}(\bar{B} \to X_s l^+ l^-)\;[10^{-6}]$ and their 
weighted averages (w.a.) for two different ranges of the dilepton invariant mass
squared:~
(a)~ $(2m_{\mu})^2 < q^2 < (m_B-m_K)^2$~
and
(b)~ $1\,{\rm GeV}^2 < q^2 < 6\,{\rm GeV}^2$.  \label{tab:exp}}
\begin{tabular}{cccc}
\hline \hline
Range & Belle~\cite{Abe:2004sg} & BaBar~\cite{Aubert:2004it} & w.a. \\
\hline
&&&\\[-2.5mm]
(a) &
$4.11 \pm 0.83 ~{}^{+0.74}_{-0.70}$ &
$5.6  \pm 1.5  \pm 0.6 \pm 1.1$     & 
$4.5 \pm 1.0$ \\[1mm]
(b) &
$1.493 \pm 0.504 ~{}^{+0.382}_{-0.283}$ &
$1.8   \pm 0.7 \pm 0.5$ & 
$1.60 \pm 0.51$ \\[0.5mm]
\hline \hline
\end{tabular}
\end{table}

The Standard Model perturbative calculations are available at the
Next-to-Next-to-Leading Order (NNLO) in QCD for both the considered ranges of
$q^2$ --- see Refs.~\cite{Bobeth:2003at,Ghinculov:2003qd} for the most recent
phenomenological analyses and a list of relevant references. The dominant
electroweak corrections are also known \cite{Bobeth:2003at}.  In the low-$q^2$
domain, non-perturbative effects are taken into account in the framework of
the Heavy Quark Expansion as $\Lambda^2/m_b^2$ and $\Lambda^2/m_c^2$
corrections \cite{nonpert}. Analytical expressions for such corrections are
also available for the full $q^2$ range, but they blow up in the vicinity of
the intermediate $\psi$ peak. Consequently, a cut needs to be applied, and it
is no longer clear what theoretical procedure corresponds to the interpolation
that is performed on the experimental side.  Thus, the relative theoretical
uncertainty for the full $q^2$ range is larger than for the low-$q^2$ window.
\begin{table}[h]
\caption{\sf 
Predictions for ${\cal B}(\bar{B} \to X_s l^+ l^-)\;[10^{-6}]$ 
in the Standard Model and with reversed sign of $\widetilde{C}_7^{\rm eff}$
for the same ranges of $q^2$ as in Tab.~\ref{tab:exp}. \label{tab:th}}
\begin{tabular}{ccccc}
\hline \hline
&&\\[-3mm]
Range && SM && $\widetilde{C}_7^{\rm eff} \to -\widetilde{C}_7^{\rm eff}$ \\
\hline
&~~&&~~~&\\[-2.5mm]
(a) && $4.4  \pm 0.7$  && $8.8  \pm 1.0$ \\[1mm]
(b) && $1.57 \pm 0.16$ && $3.30 \pm 0.25$ \\
\hline \hline
\end{tabular}
\end{table}

The results of the SM calculations are given in the central column of
Tab.~\ref{tab:th}.  For the low-$q^2$ domain, they correspond to the ones of
Ref.~\cite{Bobeth:2003at} with updated input values $m_{t, \rm pole}=178.0\pm
4.3$~GeV~\cite{Azzi:2004rc} and ${\cal B}(\bar{B} \to X_c l \bar{\nu}) = 10.61
\pm 0.16 \pm 0.06$~\cite{Aubert:2004aw}. The dominant sources of uncertainty
are the values of the top and bottom quark masses, as well as the residual
renormalization scale dependence.  For the full $q^2$ range, we make use of
the statement in Ref.~\cite{Ghinculov:2003qd} that the NNLO matrix elements
for $\hat s= q^2/m_b^2 > 0.25$ are accurately reproduced by setting the
renormalization scale $\mu_b = m_b/2$ at the Next-to-Leading Order (NLO) level.

To a very good approximation, the amplitude \linebreak
$A(b\to s \gamma)$ 
is proportional to the effective Wilson coefficient 
$\widetilde{C}_{7}^{\rm eff}(q^2=0)$ 
that determines the strength of the 
$\bar{s}_L \sigma^{\alpha\beta} b_R F_{\alpha\beta}$ 
interaction term in the low-energy Hamiltonian.  The sign of
$A(b\to s \gamma)$ 
is therefore given by the sign of 
$\widetilde{C}_{7}^{\rm eff}(q^2=0)$. 
Both the value and the sign of this coefficient matter for the
rare semileptonic decay. The results in the right column of Tab.~\ref{tab:th}
differ from those in the central column only by reversing the sign of
$\widetilde{C}_{7}^{\rm eff}$ in the expression for the differential $\bar{B}
\to X_s l^+ l^-$ decay rate
\begin{eqnarray}
\frac{d \Gamma[\bar{B} \to X_s l^+ l^-]}{d {\hat s}} & = &
\frac{G_F^2 m_{b, {\rm pole}}^5 \left | V_{ts}^* V_{tb} \right |^2}{48 \pi^3}
  \left ( \frac{\alpha_{em}}{4 \pi}\right )^2 \times 
\nonumber\\ && \hspace{-2cm} 
\times (1 - {\hat s})^2 \, \Bigg \{ (1 + 2 {\hat s}) 
\left(   \Big | \widetilde{C}_{ 9}^{\rm eff} \Big |^2
       + \Big | \widetilde{C}_{10}^{\rm eff} \Big |^2 \right) 
\nonumber \\ \label{dgds} && \hspace{-2cm} 
+ \left( 4 + \frac{8}{{\hat s}} \right ) \Big | \widetilde{C}_{7}^{\rm eff} \Big |^2  
+ 12 {\, \rm Re} \left ( \widetilde{C}_{7}^{\rm eff} 
                         \widetilde{C}_{9}^{\rm eff *} \right ) \Bigg \} \, ,  
\end{eqnarray}
where $\widetilde{C}_9^{\rm eff}$ and $\widetilde{C}_{10}^{\rm eff}$ correspond
to the low-energy interaction terms
$(\bar{s}_L \gamma_\alpha b_L) (\bar{l} \gamma^\alpha l)$
and 
$(\bar{s}_L \gamma_\alpha b_L) (\bar{l} \gamma^\alpha \gamma_5 l)$,
respectively. The definitions of all the relevant effective coefficients can
be found in Sec.~5 of Ref.~\cite{Asatrian:2001de}. We stress that
$\widetilde{C}_i^{\rm eff}$ depend on $q^2$ and do {\em not} depend on the
renormalization scale, up to residual higher-order effects. For simplicity,
some of the NNLO QCD, electroweak and non-perturbative corrections are omitted
in Eq.~(\ref{dgds}). However, all those corrections are taken into account in
our numerical results and plots.

The sensitivity of ${\cal B}(\bar{B} \to X_s l^+ l^-)$ to the sign of
$\widetilde{C}_7^{\rm eff}$ is quite pronounced because the last term in
Eq.~(\ref{dgds}) is sizeable and it interferes destructively (in the SM) with
the remaining ones.  One can see that the experimental values of the $\bar{B}
\to X_s l^+ l^-$ branching ratio in Tab.~\ref{tab:exp} differ from the values
in the right column of Tab.~\ref{tab:th} by 3$\sigma$ in both the low-$q^2$
window and the full $q^2$ range.
This fact disfavors extensions of the SM
in which the sign of $\widetilde{C}_{7}^{\rm eff}$ gets reversed while
$\widetilde{C}_9^{\rm eff}$ and $\widetilde{C}_{10}^{\rm eff}$ receive small
non-standard corrections only.
\begin{figure}[t]
\scalebox{0.85}{\includegraphics{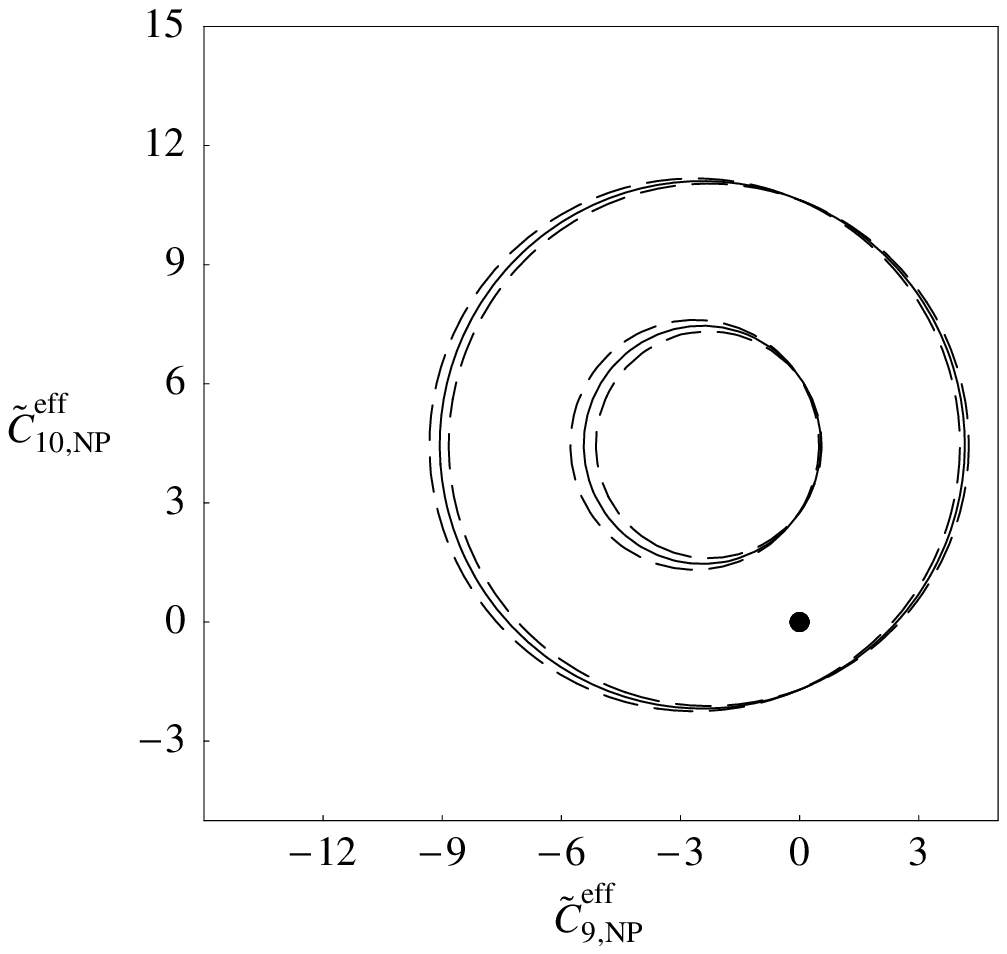}}\\[-5mm]
\scalebox{0.85}{\includegraphics{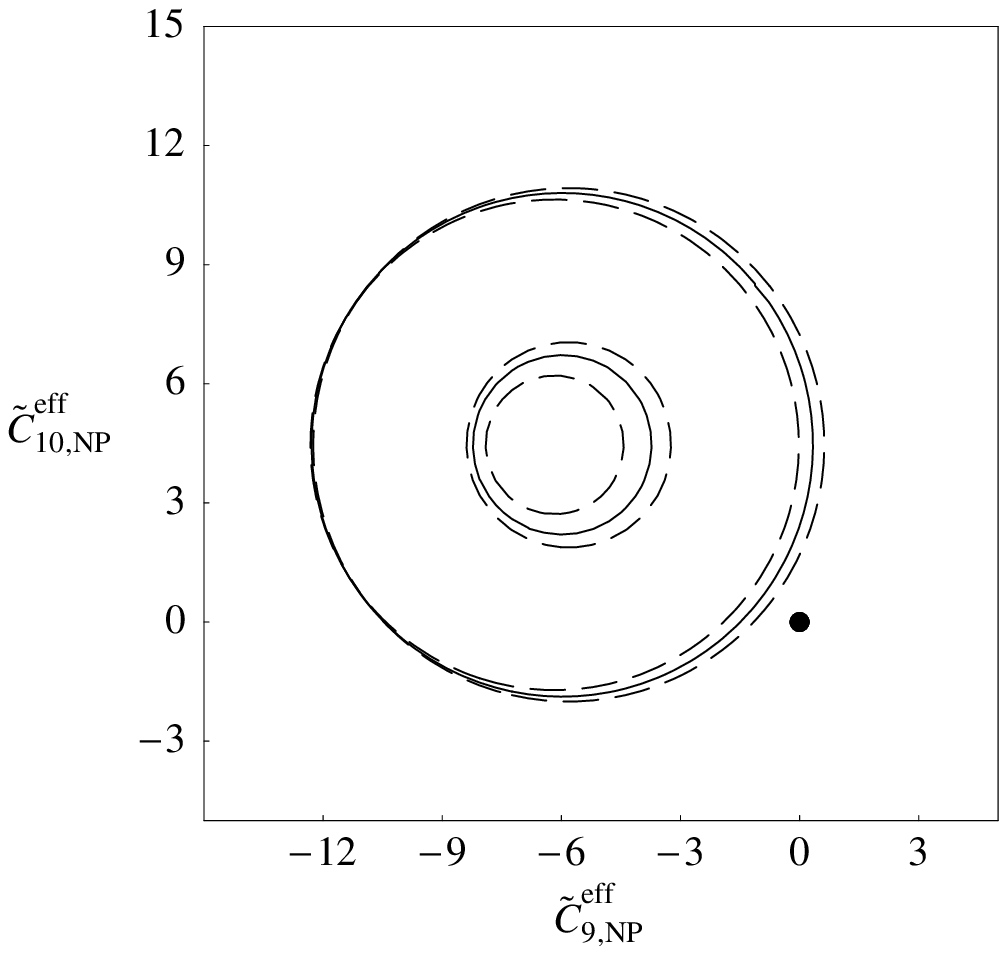}}\\[-5mm]
\caption{\sf Model-independent constraints on additive new physics
  contributions to $\widetilde{C}_{9,10}^{\rm eff}$ at 90\%~C.L. for the
  SM-like (upper plot) and the opposite (lower plot) sign of
  $\widetilde{C}_7^{\rm eff}$. The three lines correspond to three different
  values of ${\cal B}(\bar{B} \to X_s \gamma)$ (see the text).  The regions
  outside the rings are excluded. The dot at the origin indicates the SM case
  for $\widetilde{C}_{9,10}^{\rm eff}$. \label{fig:all}}
\end{figure}
\begin{figure}[t]
\begin{center}
\scalebox{0.85}{\includegraphics{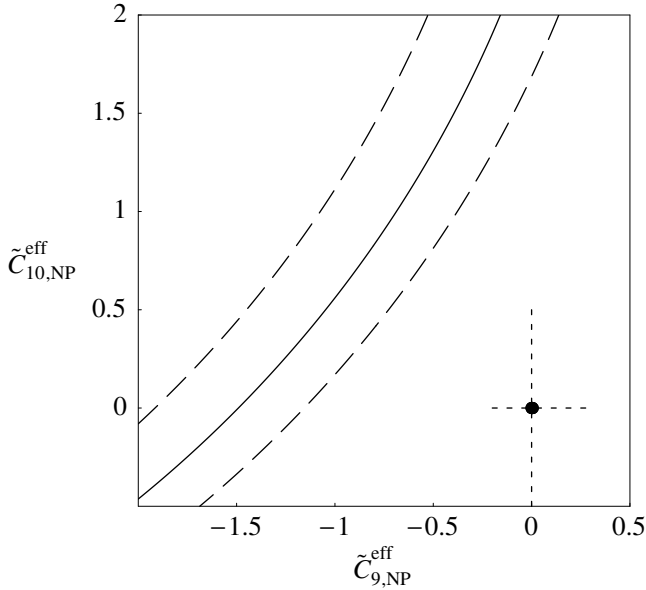}}
\end{center}
\vspace*{-9mm}
\caption{\sf Same as in the lower plot in Fig.~\ref{fig:all}. Surroundings of
  the origin. The maximal MFV MSSM ranges for $\widetilde{C}_{9,\rm NP}^{\rm
  eff}$ and $\widetilde{C}_{10,\rm NP}^{\rm eff}$ are indicated by the dashed
  cross (according to Eq.~(52) of Ref.~\cite{Ali}). \label{fig:origin}}
\end{figure}

In Fig.~\ref{fig:all}, we present constraints on additive new physics
contributions to $\widetilde{C}_{9,10}^{\rm eff}$ placed by the low-$q^2$
measurements of $\bar{B} \to X_s l^+ l^-$ (Tab.~\ref{tab:exp}), once the $\bar
B\to X_s \gamma$ bounds on $|\widetilde{C}_{7}^{\rm eff}|$ are taken into
account. Similar plots have been previously presented in
Refs.~\cite{Ali,Hiller:2003js}. The two plots correspond to the two possible
signs of the coefficient $\widetilde{C}_{7}^{\rm eff}$. The regions outside
the rings are excluded. Surroundings of the origin are magnified in
Fig.~\ref{fig:origin} for the non-standard case. The three lines correspond to
three different values of
${\cal B}(\bar B\to X_s \gamma) \times 10^4$: 2.82, 3.52 and 4.22,
which include the experimental central value as well as borders of the
90\%~C.L.  domain. In evaluating this domain, the experimental error was
enlarged by adding the SM theoretical uncertainty in quadrature. A similar
procedure was applied to ${\cal B}(\bar{B} \to X_s l^+ l^-)$. Its low-$q^2$
value was varied between $0.7\times 10^{-6}$ and $2.5 \times 10^{-6}$. The
three lines in each plot of Figs.~\ref{fig:all}~and~\ref{fig:origin} clearly
show that the exact value of $|\widetilde{C}_{7}^{\rm eff}|$ has a minor
impact on the bounds, which are therefore rather insensitive to the
theoretical error estimate in ${\cal B}(\bar B\to X_s \gamma)$.

The SM point (i.e. the origin) is located barely outside the border line of
the allowed region in the lower plot of Fig.~\ref{fig:all}. However, one
should take into account that the overall scale in this figure is huge, and
only a tiny part of the allowed region is relevant to realistic extensions of
the SM. Thus, it is more instructive to look at Fig.~\ref{fig:origin}, from
which it is evident that a non-standard sign of $\widetilde{C}_{7}^{\rm eff}$
could be made compatible with experiments only by large ${\cal O}(1)$ new
physics contributions to $\widetilde{C}_{9,10 }^{\rm eff}$. The SM values of
$\widetilde{C}_9^{\rm eff}$ and $\widetilde{C}_{10}^{\rm eff}$ are around
$+4.2$ and $-4.4$, respectively.

A case in which large non-standard contributions to $\widetilde{C}_{7}^{\rm
  eff}$ that interfere destructively with the SM ones arise naturally, while
$\widetilde{C}_{9,10}^{\rm eff}$ are only slightly affected, occurs in the
Minimal Supersymmetric Standard Model (MSSM) with Minimal Flavor Violation
(MFV) at large $\tan\beta$, with relatively light top squark and higgsino-like
chargino~\cite{Ali:1994bf,Ali,buras}. The maximal MFV MSSM contributions to
$\widetilde{C}_{9,10}^{\rm eff}$ that were found in Ref.~\cite{Ali} are
indicated by the dashed cross in Fig.~\ref{fig:origin}. As one can
see, they are too small to reach the border of the allowed region. 
For clarity, we note that although the bounds in Ref.~\cite{Ali} were given
  for the electroweak-scale Wilson coefficients, they remain practically the
  same for the $b$-scale coefficients $\widetilde{C}_{9,10}^{\rm eff}$.
  
  Configurations of the MSSM couplings and masses for which the sign of
  $\widetilde{C}_{7}^{\rm eff}$ gets reversed turn out to be relevant if no
  physics beyond the MSSM contributes to the intergalactic dark matter (see
  e.g.  Ref.~\cite{dark2}). In particular, configurations characterized by
  large mixing in the stop sector tend to be excluded by the new
  constraint~\cite{Scopel}.  While performing a dedicated scan over the MSSM
  parameters is beyond the scope of this Letter, we expect that the
  implementation of the $\bar{B} \to X_s l^+ l^-$ constraints will result in a
  significant reduction of the neutralino-dark-matter-allowed region in the
  MSSM parameter space.

One should be aware that in the MSSM at large $\tan\beta$, there are
additional contributions suppressed by powers of the lepton masses but enhanced by
$(\tan\beta)^3$. They are related to the   chirality-flip
operators 
$({\bar s}_L b_R)({\bar \mu}_L \mu_R)$
and
$({\bar s}_L b_R)({\bar \mu}_R \mu_L)$
and may contribute to the muon case in a significant way.  Fortunately, such
contributions are bounded from above~\cite{Hiller:2003js,Chankowski:2003wz} by
the experimental constraints~\cite{Acosta:2004xj} on $B^0_s \to \mu^+ \mu^- $,
and turn out to be irrelevant to
our argument.

Another interesting example occurs in the general MSSM with R-parity, where
new sources of flavor and CP violation in the squark mass matrices are
conveniently parameterized in terms of so-called mass insertions. The sign of
the $b\to s \gamma$ amplitude can be reversed without affecting
$\widetilde{C}_{9,10}^{\rm eff}$ if the mass insertion $(\delta_{23}^d)_{LR}$
is large and positive \cite{luca}. The new results on $\bar{B} \to X_s l^+ l^-$
exclude this possibility, and constrain significantly the case of a complex
$(\delta_{23}^d)_{LR}$.\\

To conclude: We have pointed out that the recent measurements of ${\cal
  B}(\bar{B} \to X_s l^+ l^-)$ by Belle and BaBar already indicate that the
sign of the $b \to s \gamma$ amplitude is unlikely to be different from that
in the SM.  This underscores the importance of $\bar{B} \to X_s l^+ l^-$ in
searches for new physics, and may be relevant for neutralino dark matter 
analyses within the MSSM.\\[5mm]
{\bf Acknowledgments}\\[-4mm]

We would like to thank I.~Blokland, C.~Bobeth, S.~Davidson, A.~Freitas,
L.~Roszkowski, S.~Scopel and L.~Silvestrini for helpful discussions and
correspondence. M.M. is grateful to INFN Torino for hospitality during his
visit. P.G.\ was supported in part by the EU grant MERG-CT-2004-511156.
U.~H.\ was supported by the U.S. Department of Energy under contract
No.~DE-AC02-76CH03000.  M.M.\ was supported in part by the Polish Committee for
Scientific Research under the grant 2~P03B~078~26, and from the European
Community's Human Potential Programme under the contract HPRN-CT-2002-00311,
EURIDICE.

 
\end{document}